\documentclass[english,final,proceeding]{jcaa}
\usepackage{subfig}
\title{EEG-to-F0: Establishing artificial neuro-muscular pathway for kinematics-based fundamental frequency control
}
\author[1,3]{Himanshu Goyal\thanks{h.goyal@alumni.ubc.ca}}
\author[2]{Pramit Saha\thanks{pramit@ece.ubc.ca}}
\author[3]{Bryan Gick\thanks{gick@mail.ubc.ca}}
\author[2]{Sidney Fels\thanks{ssfels@ece.ubc.ca}}
\affil[1]{Department of Computer Science, University of British Columbia}
\affil[2]{Department of Electrical and Computer Engineering, University of British Columbia}
\affil[3]{Department of Linguistics, University of British Columbia}
\begin{document}

\twocolumn[

\maketitle

]

\saythanks

\section{Introduction}
Speech-related Brain Computer Interfaces (BCI) are primarily targeted at finding alternative vocal communication pathways for people with speaking disabilities. However, most of the works in this field are centered around recognition of words from imagined speech, given a particular, small vocabulary \cite{saha2019speak,saha2019deep,saha2019hierarchical}. Lesser efforts have been invested towards investigating voice synthesis from the information decoded from EEG signals\cite{anumanchipalli2019speech,comstock2019developing}. In this work, as an extension of the previous work \cite{saha2019sound}, we aim at controlling the fundamental frequency of voice based on information derived from brain signals for an elbow movement task, through a biomechanical simulation toolkit ArtiSynth\cite{lloyd2012artisynth,saha2019sound}. To the best of our knowledge, this is the first attempt targeting such a brain-to-vocal frequency mapping incorporating a biomechanical control pathway. 
\section{Overview of proposed technique}
The fundamental frequency (F0) of human voice is generally controlled by changing the vocal fold parameters (including tension, length and mass), which in turn is manipulated by the muscle exciters, activated by neural synergies. In order to begin investigating the neuromuscular to F0 control pathway, we simulate a simple biomechanical arm prototype (instead of an artificial vocal tract) that tends to control F0 of an artificial sound synthesiser based on the elbow movements. The intended arm movements are decoded from the EEG signal inputs (collected simultaneously with the kinematic hand data of the participant) through a combined machine learning and biomechanical modeling strategy.  
\begin{figure}[]
\centering
\includegraphics[width=11cm,height=4.5cm,keepaspectratio]{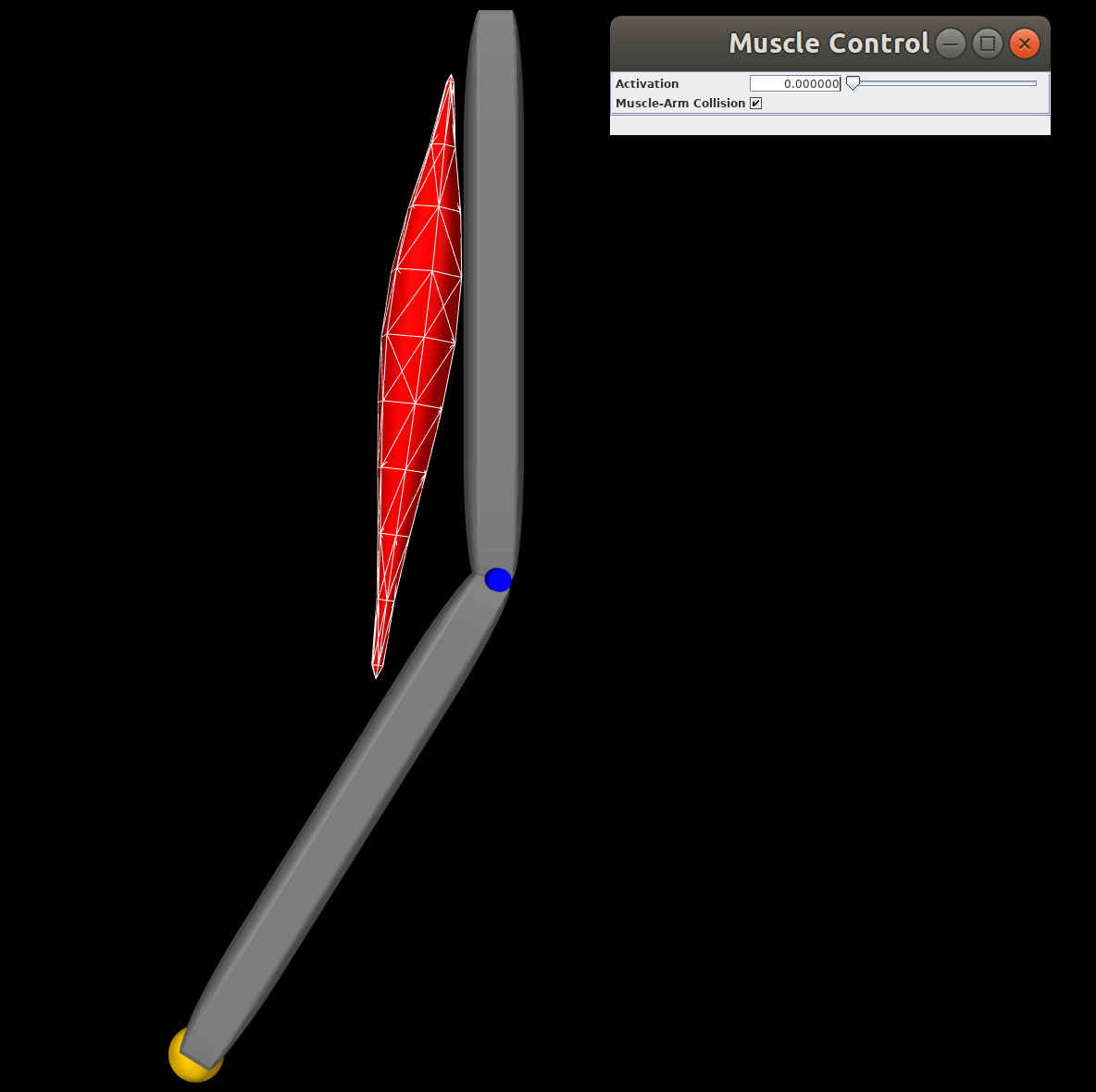}
\caption{Simulated single-muscle arm model in ArtiSynth}
\label{fig:framework}
\end{figure}
\section{Data collection and preparation}
\begin{figure}[]
\centering
\includegraphics[width=18cm,height=4.5cm,keepaspectratio]{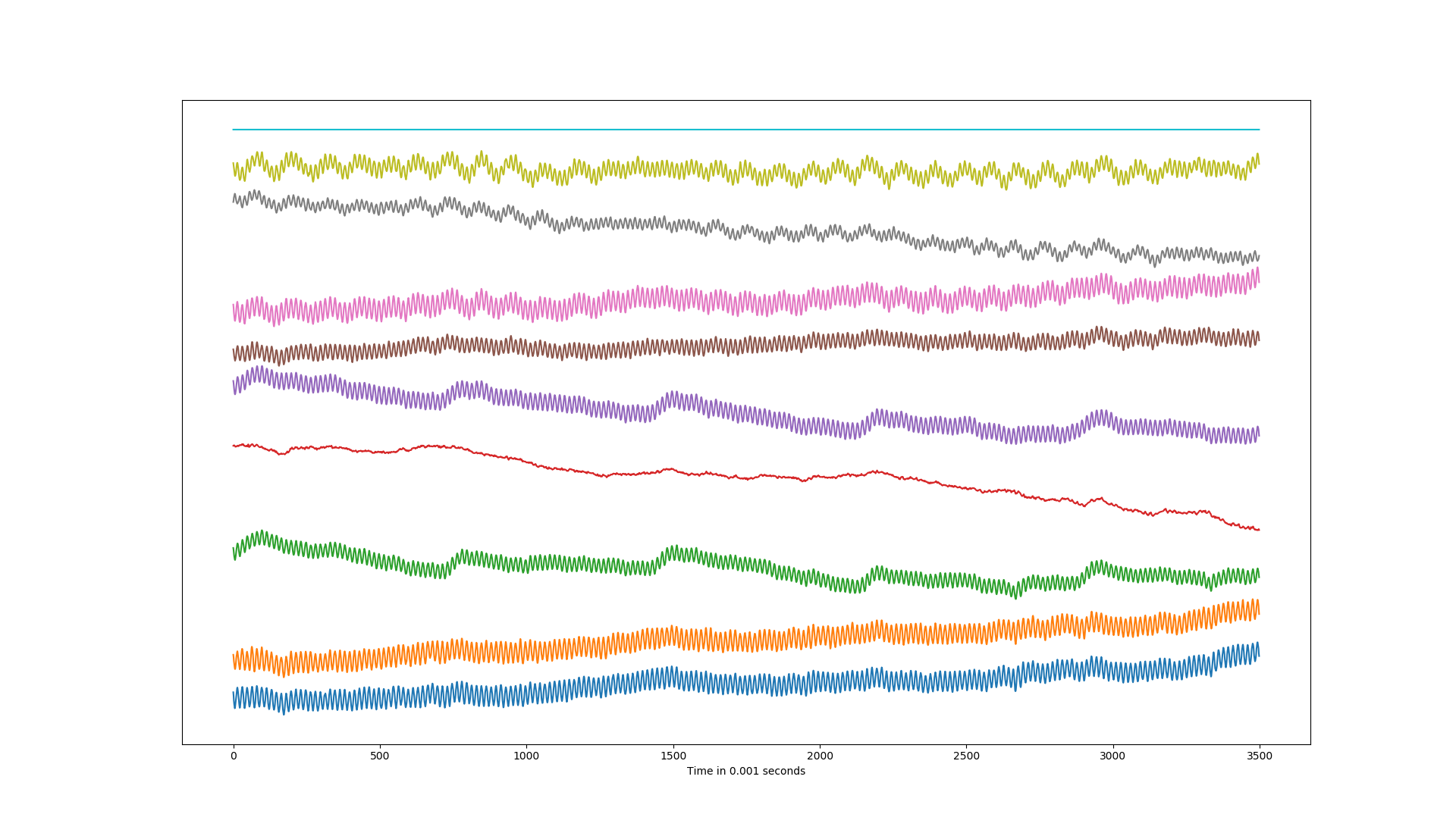}
\caption{Sample of 10 channel EEG data collected (0.001 s)}
\label{fig:framework2}
\end{figure}

We record EEG data as shown in Fig. \ref{fig:framework2}, corresponding to hand movements using a 10 electrode dry EEG headset with bluetooth connectivity, manufactured by Avertus. It has a sampling frequency of 1000 Hz per electrode at 20 bits/sample. We take data from the electrodes FP1, FP2, F7, F8, T3, T4, T5, T6, O1 and O2 with FCz as reference and FPz as the ground electrode. We particularly choose this headset because of the ease of use and portability of the equipment, which makes our strategy practically implementable.  We collect 500 EEG data samples from single participant: 250 each for the hand going up and down, both with the eyes closed and the head fixed. We do not perform any pre-processing on the data and split the data into train:test as 70:30. We also simultaneously collect the angular displacement values or the elbow angles correspondingly with the EEG. 
\section{Proposed methodology}
In order to achieve this EEG to F0 mapping, we first attempt to map the EEG signals corresponding to the elbow movements to the discretized muscle activation of the simulated one-muscle arm model (in the biomechanical toolkit ArtiSynth as shown in Fig. \ref{fig:framework}) that would create same elbow angular displacement. The temporal resolution of the muscle activations in ArtiSynth is .01 second. Hence, we set a time-window of 0.01 second for EEG signal as well, thereby, each EEG sample ( to be mapped to muscle activation) becomes a 10 x 10 matrix . Each matrix is fed to a random forest based classifier with min sample leaf = 1 and min sample split = 2, number of estimators = 10 and random state = 42. This is a 10-class classification task, with the output classes being \{0.1, 0.2, 0.3, 0.4, 0.5, 0.6, 0.7, 0.8, 0.9 and 1\} which represent the discretized muscle activations between 0 and 1. Using inverse biomechanical modeling in ArtiSynth, we calculate the required muscle activation corresponding to targeted angular movement and optimise the loss function between the intended muscle activation (from EEG signal) and simulated muscle activation (from ArtiSynth) to achieve the best possible mapping from EEG to the muscle activation.

This muscle activation is loaded into the one-muscle biomechanical model of the hand in the ArtiSynth. The resultant angular movement is recorded and the elbow joint angle is then linearly mapped to F0 between 1500 Hz and 5150 Hz. We use the F0 value mapped from the actual kinematic hand data as the ground truth and compare the F0 estimated from brain signal to evaluate the performance of our method.

\section{Results}
\begin{figure}[]
\centering
\subfloat[Actual F0 values]{{\includegraphics[width=11cm,height=4.5cm,keepaspectratio]{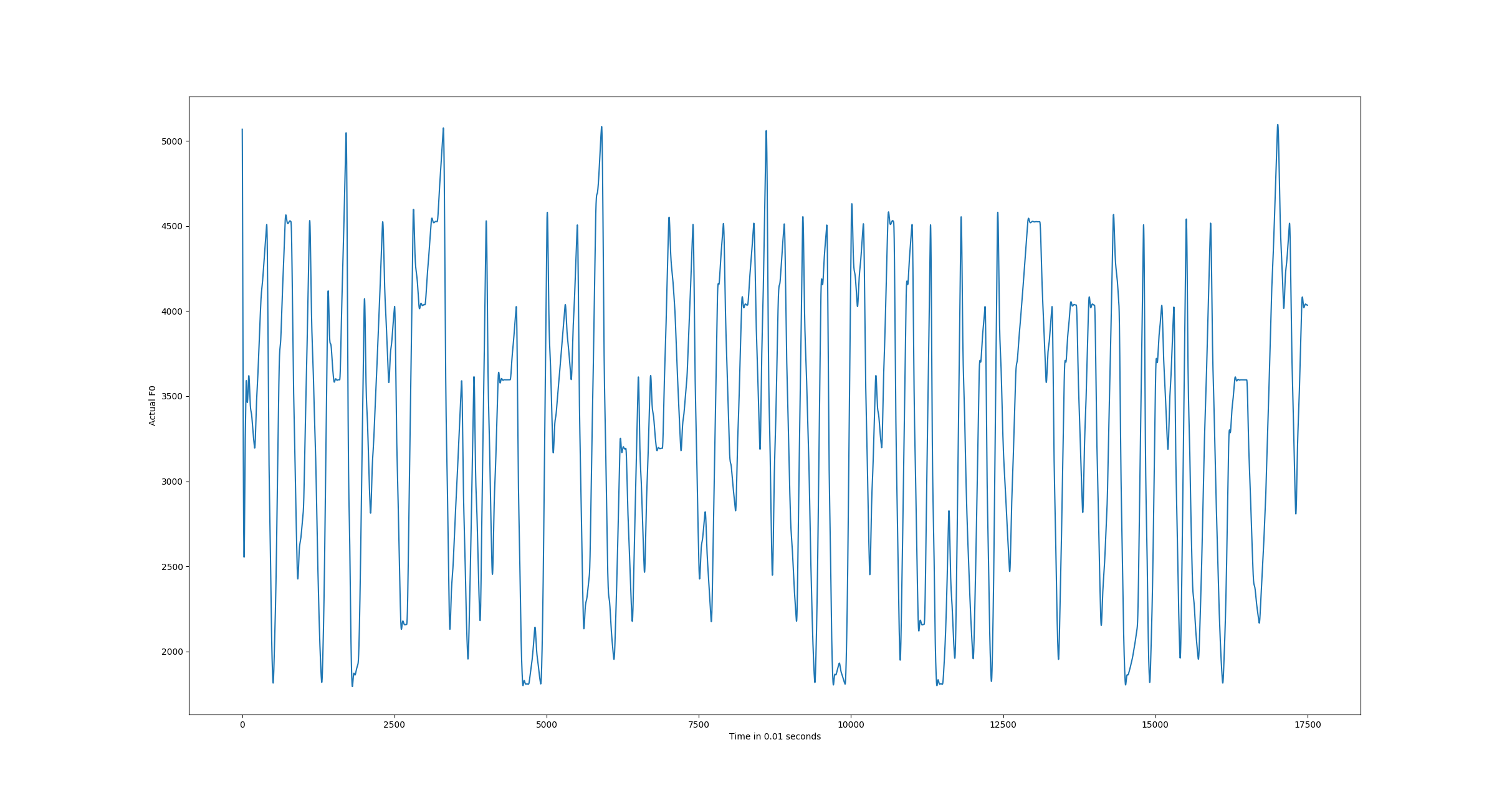} }}%
    \qquad
\subfloat[Predicted F0 values]{{\includegraphics[width=11cm,height=4.5cm,keepaspectratio]{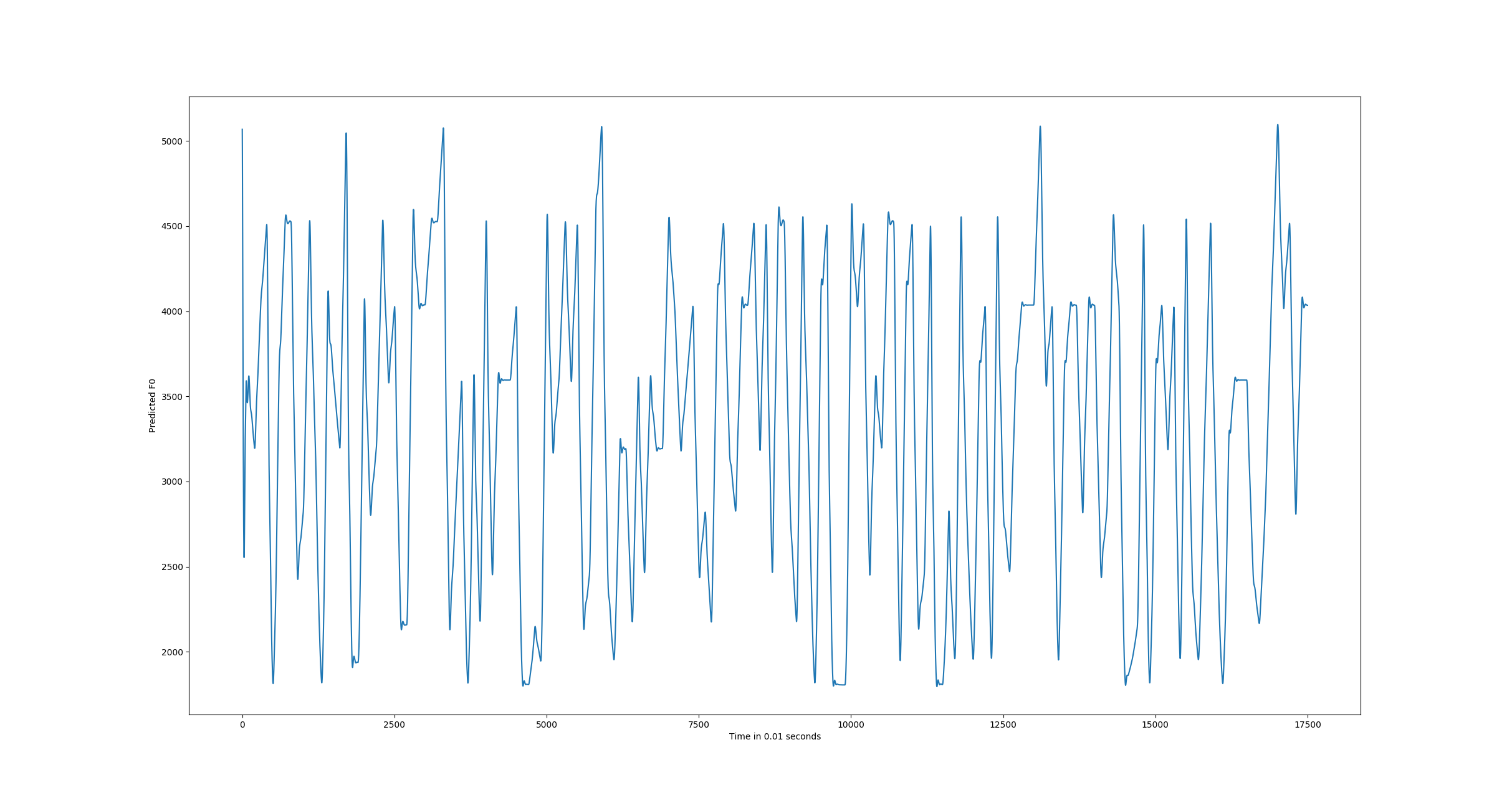} }}%
\caption{Comparison of actual and predicted (from EEG) fundamental frequency values}
\label{fig:framework1}
\end{figure}

In this section, we present our step-wise performance for the test data. In the first step, \textit{i.e.}, mapping EEG to muscle activations, the overall mean accuracy (the percentage of times, the predicted value matches the actual value) achieved is 85\%, with a root mean square error (rmse) of 0.04. After the next step, \textit{i.e.} mapping muscle activations to angles, the cumulative mean accuracy is 68.8\% with rmse of 3.4 degrees. The final rmse of the F0 estimation (ranging from 1.5 KHz to 5.15 KHz) is 102.7 Hz. For qualitative comparison of the final predicted F0 with actual F0, a randomly chosen sample is shown in Fig. \ref{fig:framework1}. 

\section{Conclusion}
We report a novel successful mapping scheme from EEG to fundamental frequency, ranging from 1.5 KHz to 5.15 KHz, with a negligible rmse of 102.7 Hz. This is an initial prototype which proved to be satisfactory in testing the idea of EEG-to-muscle activation-to-kinematics-to-acoustics mapping. Instead of using the complex biomechanical space of the vocal folds, as an initial work in this direction, we tested the concept using simple elbow rotation and single-muscle arm model to simulate the movement, where upwards (or downwards) elbow motion represented increase (or decrease) in frequency. Further work will be directed towards incorporating the ideas using multiple muscle biomechanical arm model, extending it to a vocal fold model as well as replacing the random forest-based classifier with neural network models to investigate any increase in performance accuracy. 

\section*{Acknowledgments}

This work was funded by the Natural Sciences and Engineering Research Council (NSERC) of Canada and Canadian Institutes for Health Research (CIHR). 

\bibliographystyle{unsrt}
\bibliography{references}
\nocite{*}

\end{document}